\documentclass[a4paper,12pt]{article}

\usepackage{latexsym}
\usepackage[small,hang,bf]{caption} 

\newcommand{\nc}{\newcommand}    

\nc{\be}[1]{\begin{equation}\mbox{$\label{#1}$}}
\nc{\bea}[1]{\begin{eqnarray} \mbox{$\label{#1}$}}
\nc{\Section}[2]{\section{#2}\label{#1}}
\nc{\Bibitem}[1]{\bibitem{#1}}
\nc{\Label}[1]{\label{#1}}

\nc{\eea}{\end{eqnarray}}
\nc{\ee}{\end{equation}}

\nc{\bdm}{\begin{displaymath}}
\nc{\edm}{\end{displaymath}}
\nc{\dpsty}{\displaystyle}
\nc{\bc}{\begin{center}}
\nc{\ec}{\end{center}}
\nc{\ba}{\begin{array}}
\nc{\ea}{\end{array}}
\nc{\bab}{\begin{abstract}}
\nc{\eab}{\end{abstract}}
\nc{\btab}{\begin{tabular}}
\nc{\etab}{\end{tabular}}
\nc{\bit}{\begin{itemize}}
\nc{\eit}{\end{itemize}}
\nc{\ben}{\begin{enumerate}}
\nc{\een}{\end{enumerate}}
\nc{\bfig}{\begin{figure}}
\nc{\efig}{\end{figure}}

\nc{\arreq}{&\!=\!&}
\nc{\arrmi}{&\!-\!&}
\nc{\arrpl}{&\!+\!&}
\nc{\arrap}{&\!\!\!\approx\!\!\!&}
\nc{\non}{\nonumber}
\nc{\align}{\!\!\!\!\!\!\!\!&&}

\def\lsim{\; \raise0.3ex\hbox{$<$\kern-0.75em
      \raise-1.1ex\hbox{$\sim$}}\; }
\def\gsim{\; \raise0.3ex\hbox{$>$\kern-0.75em
      \raise-1.1ex\hbox{$\sim$}}\; }

\nc{\DOT}{\hspace{-0.08in}{\bf .}\hspace{0.1in}}
\nc{\Laada}{\hbox {$\sqcap$ \kern -1em $\sqcup$}}
\nc\loota{{\scriptstyle\sqcap\kern-0.55em\hbox{$\scriptstyle\sqcup$}}}
\nc\Loota{{\sqcap\kern-0.65em\hbox{$\sqcup$}}}
\nc\laada{\Loota}
\nc{\qed}{\hskip 3em \hbox{\BOX} \vskip 2ex}

\nc{\real}{{\rm I \! R}}
\nc{\ireal}{{\it I \!\! R}}
\nc{\Z}{{\sf Z \!\!\! Z}}
\nc{\complex}{{\rm C\!\!\! {\sf I}\,\,}}
\def\bigid{\leavevmode\hbox{\small1\kern-3.8pt\normalsize1}}
\def\id{\leavevmode\hbox{\small1\kern-3.3pt\normalsize1}}
\nc{\slask}{\!\!\!/}
\nc{\bis}{{\prime\prime}}
\nc{\pa}{\partial}
\nc{\na}{\nabla}
\nc{\ra}{\rangle}
\nc{\la}{\langle}
\nc{\goto}{\rightarrow}
\nc{\swap}{\leftrightarrow}

\nc{\EE}[1]{ \mbox{$\cdot10^{#1}$} }
\nc{\abs}[1]{\left|#1\right|}
\nc{\at}[2]{\left.#1\right|_{#2}}
\nc{\norm}[1]{\|#1\|}
\nc{\abscut}[2]{\Abs{#1}_{\scriptscriptstyle#2}}
\nc{\vek}[1]{{\rm\bf #1}}
\nc{\integral}[2]{\int\limits_{#1}^{#2}}
\nc{\inv}[1]{\frac{1}{#1}}
\nc{\dd}[2]{{{\partial #1}\over{\partial #2}}}
\nc{\ddd}[2]{{{{\partial}^2 #1}\over{\partial {#2}^2}}}
\nc{\dddd}[3]{{{{\partial}^2 #1}\over
    {\partial #2 \partial #3}}}
\nc{\dder}[2]{{{d #1}\over{d #2}}}
\nc{\ddder}[2]{{{d^2 #1}\over{d {#2}^2}}}
\nc{\dddder}[3]{{d^2 #1}\over
    {d #2 d #3}}
\nc{\dx}[1]{d\,^{#1}x}
\nc{\dy}[1]{d\,^{#1}y}
\nc{\dz}[1]{d\,^{#1}z}
\nc{\dl}[1]{\frac{d\,^{#1}l}{(2\pi)^{#1}}}
\nc{\dk}[1]{\frac{d\,^{#1}k}{(2\pi)^{#1}}}
\nc{\dq}[1]{\frac{d\,^{#1}q}{(2\pi)^{#1}}}

\nc{\bfT}{{\bf T }}
\def\GeV{{\rm GeV}}

\def\TeV{{\rm\ TeV}}

\nc{\cA}{{\cal A}}
\nc{\cB}{{\cal B}}
\nc{\cD}{{\cal D}}
\nc{\cE}{{\cal E}}
\nc{\cG}{{\cal G}}
\nc{\cH}{{\cal H}}
\nc{\cL}{{\cal L}}
\nc{\cO}{{\cal O}}
\nc{\cT}{{\cal T}}
\nc{\cN}{{\cal N}}
\nc{\cR}{{\cal R}}
\nc{\rvac}[1]{|{\cal O}#1\rangle}
\nc{\lvac}[1]{\langle{\cal O}#1|}
\nc{\rvacb}[1]{|{\cal O}_\beta #1\rangle}
\nc{\lvacb}[1]{\langle{\cal O}_\beta #1 |}
\nc{\bb}{\bar{\beta}}
\nc{\bt}{\tilde{\beta}}
\nc{\ctH}{\tilde{\cal H}}
\nc{\chH}{\hat{\cal H}}
\nc{\1}{\aa}
\nc{\2}{\"{a}}
\nc{\3}{\"{o}}
\nc{\4}{\AA}
\nc{\5}{\"{A}}
\nc{\6}{\"{O}}
\nc{\al}{\alpha}
\nc{\g}{\gamma}
\nc{\Del}{\Delta}
\nc{\e}{\textrm{e}}
\nc{\eps}{\epsilon}
\nc{\lam}{\lambda}
\nc{\Om}{\Omega}
\nc{\ve}{\varepsilon}
\nc{\mn}{{\mu\nu}}
\nc{\vp}{\varphi}

\nc{\advp}[3]{{\it  Adv.\ in\ Phys.\ }{{\bf #1} {(#2)} {#3}}}
\nc{\annp}[3]{{\it  Ann.\ Phys.\ (N.Y.)\ }{{\bf #1} {(#2)} {#3}}}
\nc{\apl}[3]{{\it  Appl. Phys. Lett. }{{\bf #1} {(#2)} {#3}}}
\nc{\apj}[3]{{\it  Ap.\ J.\ }{{\bf #1} {(#2)} {#3}}}
\nc{\apjl}[3]{{\it  Ap.\ J.\ Lett.\ }{{\bf #1} {(#2)} {#3}}}
\nc{\app}[3]{{\it Astropart.\ Phys.\ }{{\bf #1} {(#2)} {#3}}}
\nc{\cmp}[3]{{\it  Comm.\ Math.\ Phys.\ }{{ \bf #1} {(#2)} {#3}}}
\nc{\cqg}[3]{{\it  Class.\ Quant.\ Grav.\ }{{\bf #1} {(#2)} {#3}}}
\nc{\epl}[3]{{\it  Europhys.\ Lett.\ }{{\bf #1} {(#2)} {#3}}}
\nc{\ijmp}[3]{{\it Int.\ J.\ Mod.\ Phys.\ }{{\bf #1} {(#2)} {#3}}}
\nc{\ijtp}[3]{{\it Int.\ J.\ Theor.\ Phys.\ }{{\bf #1} {(#2)} {#3}}}
\nc{\jhep}[3]{{\it  JHEP\ }{{ \bf #1} {(#2)} {#3}}}
\nc{\jetpl}[3]{{\it  JETP Lett.\ }{{ \bf #1} {(#2)} {#3}}}
\nc{\jmp}[3]{{\it  J.\ Math.\ Phys.\ }{{ \bf #1} {(#2)} {#3}}}
\nc{\jpa}[3]{{\it  J.\ Phys.\ A\ }{{\bf #1} {(#2)} {#3}}}
\nc{\jpc}[3]{{\it  J.\ Phys.\ C\ }{{\bf #1} {(#2)} {#3}}}
\nc{\jap}[3]{{\it J.\ Appl.\ Phys.\ }{{\bf #1} {(#2)} {#3}}}
\nc{\jpsj}[3]{{\it J.\ Phys.\ Soc.\ Japan\ }{{\bf #1} {(#2)} {#3}}}
\nc{\lmp}[3]{{\it Lett.\ Math.\ Phys.\ }{{\bf #1} {(#2)} {#3}}}
\nc{\mpl}[3]{{\it  Mod.\ Phys.\ Lett.\ }{{\bf #1} {(#2)} {#3}}}
\nc{\ncim}[3]{{\it  Nuov.\ Cim.\ }{{\bf #1} {(#2)} {#3}}}
\nc{\np}[3]{{\it  Nucl.\ Phys.\ }{{\bf #1} {(#2)} {#3}}}
\nc{\pr}[3]{{\it Phys.\ Rev.\ }{{\bf #1} {(#2)} {#3}}}
\nc{\pra}[3]{{\it  Phys.\ Rev.\ A\ }{{\bf #1} {(#2)} {#3}}}
\nc{\prb}[3]{{\it  Phys.\ Rev.\ B\ }{{{\bf #1} {(#2)} {#3}}}}
\nc{\prc}[3]{{\it  Phys.\ Rev.\ C\ }{{\bf #1} {(#2)} {#3}}}
\nc{\prd}[3]{{\it  Phys.\ Rev.\ D\ }{{\bf #1} {(#2)} {#3}}}
\nc{\prl}[3]{{\it Phys.\ Rev.\ Lett.\ }{{\bf #1} {(#2)} {#3}}}
\nc{\pl}[3]{{\it  Phys.\ Lett.\ }{{\bf #1} {(#2)} {#3}}}
\nc{\prep}[3]{{\it Phys\. Rep.\ }{{\bf #1} {(#2)} {#3}}}
\nc{\prsl}[3]{{\it Proc.\ R.\ Soc.\ London\ }{{\bf #1} {(#2)} {#3}}}
\nc{\ptp}[3]{{\it  Prog.\ Theor.\ Phys.\ }{{\bf #1} {(#2)} {#3}}}
\nc{\ptps}[3]{{\it  Prog\ Theor.\ Phys.\ suppl.\ }{{\bf #1} {(#2)} {#3}}}
\nc{\physa}[3]{{\it  Physica\ A\ }{{\bf #1} {(#2)} {#3}}}
\nc{\physb}[3]{{\it  Physica\ B\ }{{\bf #1} {(#2)} {#3}}}
\nc{\phys}[3]{{\it Physica\ }{{\bf #1} {(#2)} {#3}}}
\nc{\rmp}[3]{{\it  Rev.\ Mod.\ Phys.\ }{{\bf #1} {(#2)} {#3}}}
\nc{\rpp}[3]{{\it Rep.\ Prog.\ Phys.\ }{{\bf #1} {(#2)} {#3}}}
\nc{\sjnp}[3]{{\it Sov.\ J.\ Nucl.\ Phys.\ }{{\bf #1} {(#2)} {#3}}}
\nc{\spjetp}[3]{{\it Sov.\ Phys.\ JETP\ }{{\bf #1} {(#2)} {#3}}}
\nc{\yf}[3]{{\it Yad.\ Fiz.\ }{{\bf #1} {(#2)} {#3}}}
\nc{\zetp}[3]{{\it Zh.\ Eksp.\ Teor.\ Fiz.\  }{{\bf #1}  {(#2)} {#3}}}
\nc{\zp}[3]{{\it Z.\ Phys.\ }{{\bf #1} {(#2)} {#3}}}
\nc{\ibid}[3]{{\sl ibid.\ }{{\bf #1} {#2} {#3}}}

\nc{\rf}[1]{(\ref{#1})}
\nc{\nn}{\nonumber \\*}
\nc{\bfB}{\bf{B}}
\nc{\bfv}{\bf{v}}
\nc{\bfx}{\bf{x}}
\nc{\bfy}{\bf{y}}
\nc{\vx}{\vec{x}}
\nc{\vy}{\vec{y}}
\nc{\oB}{\overline{B}}
\nc{\oI}{\overline{I}}
\nc{\oR}{\overline{R}}
\nc{\rar}{\rightarrow}
\nc{\ti}{\times}
\nc{\slsh}{\hskip-5pt/}
\nc{\sm}{Standard~Model~}
\nc{\MP}{M_{\rm Pl}}
\nc{\tp}{t_{\rm Pl}}

\nc{\pmin}{p_{\rm min}}
\nc{\pmax}{p_{\rm max}}
\nc{\fo}{f_0}
\nc{\foi}{f_{0,i}\,}
\nc{\fop}{f_0^P}
\nc{\fou}{f_0^U}

\nc{\eff}{{\rm eff}}
\nc{\MT}{M_{\rm T}}
\nc{\ML}{M_{\rm L}}
\nc{\kk}{\vek{k}}
\nc{\pp}{{\rm p}}
\nc{\pt}{\partial_t}
\nc{\half}{{1\over 2}}
\nc{\w}{\omega}
\nc{\uhat}{\hat{U}_\w}

\nc{\etal}{\mbox{\it et al.}}
\nc{\ie}{{\it i.e. }}
\nc{\eg}{{\it e.g. }}
\nc{\trh}{T_{\rm RH}}
\nc{\ad}{{a'\over a}}
\nc{\bd}{{b'\over b}}
\nc{\Rd}{{R'\over R}}
\nc{\diag}{{\textrm{diag}}}
\nc{\mato}[1]{\tilde{#1}}
\nc{\sech}{\textrm{sech}}
\nc{\I}{\textrm{I}}
\nc{\II}{\textrm{II}}
\nc{\III}{\textrm{III}}

\begin{document}
\title{{\hfill {{\small  TURKU-FL-P41-02
        }}\vskip 1truecm}
{\LARGE 
Warped and compact extra dimensions:\\
5D branes in 6D models}
\vspace{-.2cm}}

\author{{\sc Tuomas Multam\"aki\footnote{email: tuomul@utu.fi}}\\ 
and\\
{\sc Iiro Vilja\footnote{email: vilja@utu.fi}}\\
\\{\sl\small Department of Physics, University of Turku, FIN-20014, FINLAND}}

\date{}
\maketitle
\thispagestyle{empty} 

\abstract{We consider six dimensional brane world models with a compact and
a warped extra dimension with five dimensional branes. We find that 
such scenarios have many interesting features arising from both 
ADD and Randall-Sundrum -models. In particular we study a class of models
with a single 5D brane and a finite warped extra dimension, where one of 
the brane dimensions is compact. In these models the hierarchy problem 
can be solved on a single positive tension brane.
}

\setlength{\captionmargin}{30pt}

\newpage
\setcounter{page}{1} 

\section{Introduction}

Extra dimensions, both compact (ADD) \cite{add} and warped (RS) \cite{rs}, 
have received an immense amount of attention in recent years \cite{rubakov}. 
The theoretical motivation coming from string theory
and the realizations that gravity is not experimentally 
well known at smaller than millimeter scales have inspired much 
of the research effort. Furthermore the properties of warped 
extra dimensions that help to alleviate the hierarchy problem with
an experimentally reachable gravity scale are of great interest.

The majority of the papers on the subject of warped
extra dimensions have concentrated on the Randall-Sundrum
-scenarios, with one or more branes, in five dimensions. 
More recently, a number of six dimensional constructions
have been considered as well (see \eg \cite{misha}-\cite{chacko}).
In the case of compact surplus spaces, there is a large number of studies
from one up to several dimensions. To solve the hierarchy problem, the
ADD-case requires, however, at least two compact dimensions.

In this paper we are interested in the properties of 6D models
that combine both the compact and warped geometries.
In general, however, we are considering extra dimensions with a
non-trivial topology, \ie the topology can be more complicated
than the simple direct product, $\textrm{S}^1\times\real$. This 
allows us to include new properties to the model. Indeed, our 
scenario has no 3-brane but instead 
a (5-dimensional) 4-brane surrounded by 6-dimensional space. 
This setting puts then physical constraints on the extra brane 
dimension which has to be compact. 

The plan of the paper is as follows: in Section 2 we write down the
metric and the corresponding Einstein's equations. Static solutions of
the Einstein's equations are then studied in empty space.
Branes are added to the picture in Section 3. In Section 4 we consider
gravitons to determine the condition for the zero-mode localization 
and write the corrections to Newton's law. In Section 5 we study more carefully
a particular single brane model with finite extra dimensions.
Cosmological aspects of 6D scenarios are considered in 
Section 6. The conclusions have been drawn in Section 7.

\section{Einstein's Equations}

The metric that includes a compact and a warped dimension, can be written in the
form
\be{metric}
ds^2=\eta(\tau,z)^2d\tau^2-R(\tau,z)^2\delta_{ij}dx^idx^j-a(\tau,z)^2dz^2-
b(\tau,z)^2d\theta^2.
\ee
The $z$-coordinate corresponds to the warped direction and the
$\theta$-coordinate to the compact dimension. 
Einstein's equations read
\be{etens}
G_{AB}\equiv R_{AB}-{1\over 2}g_{AB}{\cal R}=-8\pi G_6T_{AB}-\Lambda_{AB},
\ee
where $G_6$ is the 6D Newton's constant, $T_{AB}$ the energy-impulse
tensor and the components of the inhomogeneous 6D cosmological constant
$\Lambda$, $\Lambda_z$ and $\Lambda_{\theta}$
are included in
\be{ihlambda}
\Lambda_{A}^B=
\diag(\Lambda,\Lambda,\Lambda,\Lambda,\Lambda_z,\Lambda_{\theta}).
\ee 
Note that either or both 
$\Lambda_z$ and $\Lambda_{\theta}$ may be unequal to $\Lambda$ because
there are more degrees of freedom in the metric \cite{ross,casi}.

The non-zero components of the Einstein's tensor in this general 
case can be computed in a straightforward manner:
\bea{cosmoees}
G_{00} & = & - \frac{{{\eta}}^2\,a'\,b'}
     {{a}^3\,b}  - 
  \frac{3\,{{\eta}}^2\,a'\,R'}
   {{a}^3\,R} + \frac{3\,{{\eta}}^2\,b'\,
     R'}{{a}^2\,b\,R} + 
  \frac{3\,{{\eta}}^2\,{R'}^2}
   {{a}^2\,{R}^2} + \frac{{{\eta}}^2\,b''}
   {{a}^2\,b} +\\
& & \frac{3\,{{\eta}}^2\,R''}
   {{a}^2\,R} - \frac{\dot{a}\,\dot{b}}
   {a\,b} - \frac{3\,\dot{a}\,\dot{R}}{a\,R} - 
  \frac{3\,\dot{b}\,\dot{R}}{b\,R} - 
  \frac{3\,{\dot{R}}^2}{{R}^2}\\
G_{11} & = & \frac{{R}^2\,a'\,b'}{{a}^3\,b} + 
  \frac{{R}^2\,a'\,{\eta}'}
   {{a}^3\,{\eta}} - 
  \frac{{R}^2\,b'\,{\eta}'}
   {{a}^2\,b\,{\eta}} + 
  \frac{2\,R\,a'\,R'}{{a}^3} - 
  \frac{2\,R\,b'\,R'}{{a}^2\,b} - 
  \frac{2\,R\,{\eta}'\,R'}
   {{a}^2\,{\eta}} - 
  \frac{{R'}^2}{{a}^2} -\\ 
& & \frac{{R}^2\,b''}{{a}^2\,b} - 
  \frac{{R}^2\,{\eta}''}
   {{a}^2\,{\eta}} - 
  \frac{2\,R\,R''}{{a}^2} + 
  \frac{{R}^2\,\dot{a}\,\dot{b}}
   {a\,b\,{{\eta}}^2} - 
  \frac{{R}^2\,\dot{a}\,\dot{\eta}}
   {a\,{{\eta}}^3} -
  \frac{{R}^2\,\dot{b}\,\dot{\eta}}
   {b\,{{\eta}}^3} +\\ 
& &\frac{2\,R\,\dot{a}\,\dot{R}}
   {a\,{{\eta}}^2} + 
  \frac{2\,R\,\dot{b}\,\dot{R}}
   {b\,{{\eta}}^2} - 
  \frac{2\,R\,\dot{\eta}\,\dot{R}}
   {{{\eta}}^3} + 
  \frac{{\dot{R}}^2}{{{\eta}}^2} + 
  \frac{{R}^2\,\ddot{a}}{a\,{{\eta}}^2} + 
  \frac{{R}^2\,\ddot{b}}{b\,{{\eta}}^2} + 
  \frac{2\,R\,\ddot{R}}{{{\eta}}^2}\\
G_{44} & = & - \frac{b'\,{\eta}'}
     {b\,{\eta}}  - 
  \frac{3\,b'\,R'}{b\,R} - 
  \frac{3\,{\eta}'\,R'}
   {{\eta}\,R} - 
  \frac{3\,{R'}^2}{{R}^2} - 
  \frac{{a}^2\,\dot{b}\,\dot{\eta}}
   {b\,{{\eta}}^3} + 
  \frac{3\,{a}^2\,\dot{b}\,\dot{R}}
   {b\,{{\eta}}^2\,R} -\\ 
& &  \frac{3\,{a}^2\,\dot{\eta}\,\dot{R}}
   {{{\eta}}^3\,R} + 
  \frac{3\,{a}^2\,{\dot{R}}^2}
   {{{\eta}}^2\,{R}^2} + 
  \frac{{a}^2\,\ddot{b}}{b\,{{\eta}}^2} + 
  \frac{3\,{a}^2\,\ddot{R}}{{{\eta}}^2\,R}\\
G_{55} & = & \frac{{b}^2\,a'\,{\eta}'}
   {{a}^3\,{\eta}} + 
  \frac{3\,{b}^2\,a'\,R'}{{a}^3\,R} - 
  \frac{3\,{b}^2\,{\eta}'\,R'}
   {{a}^2\,{\eta}\,R} - 
  \frac{3\,{b}^2\,{R'}^2}{{a}^2\,{R}^2} - 
  \frac{{b}^2\,{\eta}''}
   {{a}^2\,{\eta}} - 
  \frac{3\,{b}^2\,R''}{{a}^2\,R} -\\ 
& & \frac{{b}^2\,\dot{a}\,\dot{\eta}}
   {a\,{{\eta}}^3} + 
  \frac{3\,{b}^2\,\dot{a}\,\dot{R}}
   {a\,{{\eta}}^2\,R} - 
  \frac{3\,{b}^2\,\dot{\eta}\,\dot{R}}
   {{{\eta}}^3\,R} + 
  \frac{3\,{b}^2\,{\dot{R}}^2}
   {{{\eta}}^2\,{R}^2} + 
  \frac{{b}^2\,\ddot{a}}{a\,{{\eta}}^2} + 
  \frac{3\,{b}^2\,\ddot{R}}{{{\eta}}^2\,R}\\
G_{40} & = &  \frac{b'\,\dot{a}}{a\,b}  - 
  \frac{3\,R'\,\dot{a}}{a\,R} - 
  \frac{{\eta}'\,\dot{b}}
   {b\,{\eta}} - 
  \frac{3\,{\eta}'\,\dot{R}}
   {{\eta}\,R} + \frac{\dot{b'}}{b} + 
  \frac{3\,\dot{R'}}{R},
\eea
where a dot denotes a derivative with respect to the conformal time
$\tau$ and a prime denotes a derivative with respect to the $z$-coordinate.
Note that the $(2,2)$ and $(3,3)$ components are equal to the $(1,1)$
component and are therefore omitted.

\subsection{Static solutions in empty space}

Before considering brane-world scenarios or cosmological evolution, 
we first study the possible static space-time configurations allowed by the
Einstein's equations. In the static 4D Poincare 
invariant -case ($R(z)=\eta(z)$) the metric can be written in the form
\be{metric2}
ds^2=a(z)^2\eta_{\mu\nu}dx^{\mu}dx^{\nu}-dz^2-b(z)^2d\theta^2.
\ee
The Einstein's equations in empty space ($T_{AB}=0$) are then
simplified to 
\bea{stat2}
\frac{3\,b'\,a'}{b\,a} + \frac{3\,{a'}^2}{{a}^2} + 
  \frac{b''}{b} + \frac{3\,a''}{a} & = & -\Lambda\\
\frac{4\,b'\,a'}{b\,a} + \frac{6\,{a'}^2}{{a}^2} & = & -\Lambda_z\label{ste2}\\
\frac{6\,{a'}^2}{{a}^2} + \frac{4\,a''}{a}  & = & 
-\Lambda_{\theta}\label{e3},
\eea
where the prime again indicates a derivative with respect to $z$. 
Note that the dynamical solutions, where
$\eta(\tau ) \ne R(\tau )$ are possible, and expected, in cosmological
situations where the extra dimensions undergo evolution. These will
be discussed in a later Section.

For static case, depending on the values of $\Lambda$, there is a number
of possibilities for the solutions of $a$ and $b$.
In addition to the trivial solution with vanishing cosmological
constants, a number of other solutions can also be found.
From (\ref{e3}), we find the usual exponential solution,
\be{asol1}
a(z)=a_0\e^{-k(z-z_0)},
\ee
which requires that $k^2=-\Lambda_{\theta}/10$.
The exponential solution for $a(z)$ implies that
\be{bsol1}
b(z)=b_0\e^{-l(z-z_1)},
\ee
where 
\bea{landk}
l & = & {1\over 4k}(-\Lambda_z+{3\over 5}\Lambda_{\theta}).
\eea
Eq. (\ref{stat2}) then gives a constraint for the different $\Lambda$
parameters
\be{lamcon}
-{3\over 2}\Lambda_{\theta}+4\Lambda-{5\over 2}{\Lambda_z^2\over
\Lambda_{\theta}}=0.
\ee
From (\ref{landk}) we see that $k$ and $l$ can have different
signs if $-\Lambda_z+{3\over 5}\Lambda_{\theta}<0$.

As special cases we have two simple possibilities. If $b(z)$ is 
a constant, 
the model is simply the RS-model with a compact dimension. The 
values of the cosmological constants are related then by 
$\Lambda=\Lambda_z=3\Lambda_{\theta}/5$ and $a(z)=a_0\exp(-kz)$.
If, on the other hand, $a(z)$ is constant, we must set $\Lambda_z=
\Lambda_{\theta}=0,\ l^2=-\Lambda$, with $b(z)=b_0\exp(-lz)$.

We see that different values of $\Lambda_i$ allows us to have
different types of space-time configurations. The most commonly
considered possibility (\eg see \cite{misha}) is that both $k$ and 
$l$ and are positive and hence $a$ and $b$ decrease with increasing $z$.
One can also have a situation where $k>0$ and $l<0$ which implies
that the radius of the compact dimension grows exponentially 
with $z$. An interesting possibility is also the $k<0$, $l>0$
case. 


\section{Branes}\label{branesetups}

We now add branes to the picture and consider the possible
space-time configurations.
The energy momentum tensor of a 5-brane located at $z=z_0$, 
can be of the form \cite{casi}
\be{brane}
T^A_B=\delta(z-z_0)\left(\begin{array}{ccc} \sigma\delta_{\mu}^{\nu} & & \\
& 0 & \\
& & \sigma_{\theta}
\end{array}\right).
\ee
Note that by assuming that matter branes are described by Eq. (\ref{brane}),
the SM fields, as well as gravity of course, are free to propagate 
in the compact dimension.
Experiments hence give constraints on the possible parameters
of the model. This will be discussed in more detail in Section 5.2.

From the Einstein's equations (\ref{stat2})-(\ref{e3}) wee see that
the jump conditions at the brane at $z=z_0$ are:
\bea{statjumps}
{[b']\over b}\bigg |_{z_0}+3{[a']\over a}\bigg |_{z_0} & = & 
-8\pi G_6\sigma \nonumber \\
4{[a']\over a}\bigg |_{z_0} & = & -8\pi G_6\sigma_{\theta}
\eea
These jump conditions then lead to the fine-tuning between the bulk
cosmological constant(s) and brane tension, \eg in the case where
$\Lambda=\Lambda_z=\Lambda_{\theta}$, $k=l$, and if the brane
is set at the origin, we need to set, 
$\sigma=\sigma_{\theta}=-\Lambda/(10\pi G_6 k)$.

With branes in the picture, we can construct a number
of different models with 4-branes, {\it i.e.} compact branes
located at $z>0$. Some of the possibilities are
studied in the following.

\subsection{Brane setups}

\begin{figure}[ht]
\leavevmode
\centering
\vspace*{5mm}
\includegraphics{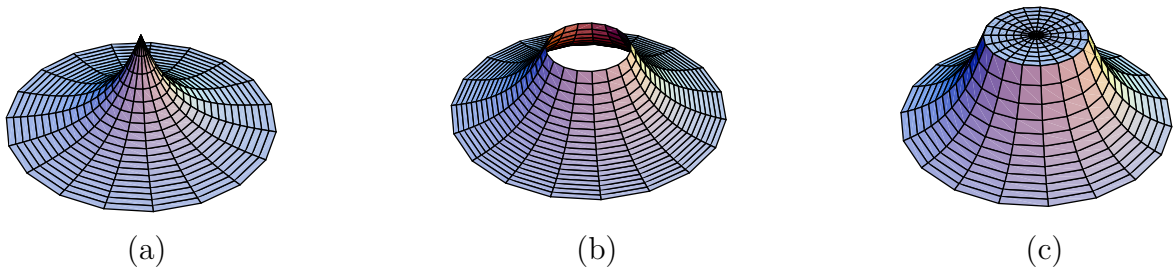}
\put(-182,-150){(a)}
\put(-12,-150){(b)}
\put(158,-150){(c)}
\caption{Some possible space-time configurations: a 4D brane at the origin (a), 
a 5D brane dislocated out of origin with no space in the middle (b) and the 
middle filled with a flat space (c). In each case there may/may not be another 
brane at the outer boundary. If no brane is placed there, 
space continues to infinity}\label{splats2}
\end{figure}

In the general case, we now have the two jump conditions (\ref{statjumps})
determined
by the brane tensions $\sigma$ and $\sigma_{\theta}$. 
One can device a number of different possible space-time configurations. 
With $k>0$, the scale factor $a$ shrinks with increasing $z$, similar to 
RS-models and one can do the obvious generalizations of the
5D RS-models to 6D.
Fig. \ref{splats2}a. is a configuration where the four-brane
is actually a string located at $z=0$. This model has been considered
in detail in \cite{misha}. We can also move the brane from 
the origin, Fig. 1b.,
and get another 6D version of the RSII-setup with 5D branes. 
Obviously, one can always add another brane along the radial dimension
which leads to a setup similar to the RSI-model.
In this case the warped direction is 
no longer infinite and one can hope to alleviate the
hierarchy problem with such a setup. Similar constructions have been 
considered in \cite{chacko}.
We can also have similar setups with $k<0$,
which may be acceptable or not, depending on the
localization of the zero-mode.
We can also combine the two solutions, so that \eg $a=a_0\exp(-k|z-z_0|)$,
and we have a volcano universe.
Note that in order to account for the jump of $a'$ at the
origin, we must also add a brane there. 

So far we have assumed that the values of $\Lambda_i$
are same constant in all regions. If we allow for different values of
$\Lambda_i$, we see that in addition to the exponential solution
there exists also the trivial solution, $a=const$, when 
$\Lambda_{\theta}=\Lambda_z=0$. One can then consider 
configurations like those depicted in Fig \ref{splats2}c.

\section{Graviton spectrum}
An important aspect of the brane world constructions is the behaviour
of gravitons. Consider a perturbation of the static 4D-Poincare invariant 
metric (\ref{metric2}),
\be{grmetric}
ds^2=a(z)^2(\eta_{\mu\nu}+h_{\mu\nu}(x,z,\theta))dx^{\mu}dx^{\nu}-
dz^2-b(z)^2d\theta^2.
\ee
Using the results derived in \cite{csaki}, the equation for linear fluctuations
of the metric is easily calculable:
\be{hequ}
h''+({b'\over b}+4{a'\over a})h'+{1\over b^2}\partial_{\theta}^2h-
{1\over a^2}\Loota h=0,
\ee
where $\Loota$ is the 4D flat space d'Alembertian operator
and prime denotes differentiation with respect to the $z$-coordinate.
Note that in deriving (\ref{hequ}), the usual fine tuning between 
the bulk cosmological constant, $\Lambda$, and brane tension, $\sigma$,
has been assumed. Hence, when considering space-times
with a different cosmological constant on different sides
of the brane, one must careful in using (\ref{hequ}) to 
calculate the graviton spectrum.

Decomposing the fluctuations, $h(z,x,\theta)=\psi(z)\varphi(x)e^{in\theta}$,
and assuming that a mode has a 4D KK-mass of $m$, (\ref{hequ}) can be
written in the form
\be{psiequ}
\psi''+({b'\over b}+4{a'\over a})\psi'+({m^2\over a^2}-{n^2\over b^2})\psi=0.
\ee
From (\ref{psiequ}) it is clear that the zero-mode, $m=0,\ n=0$,
always has a solution $\psi=const$. 
Normalizability of the zero-mode, $\int dz\sqrt{-g}g^{00} < \infty$,
hence requires that
\be{normcond}
\int dz\, a^2b<\infty.
\ee

\subsection{Localization of the graviton zero-mode}
With Eq. (\ref{normcond}) we can study the
different scenarios presented in the previous Section.
Clearly, we do not have to worry about the normalization,
and hence localization, of the graviton zero-mode unless
we have an infinite $z$-dimension. 
In the RSII-type setup with an infinite $z$-direction 
with the exponential solutions of the scale factors, (\ref{normcond}) dictates
that we must require that $2k+l>0$. In addition to the
obvious case where both $a$ and $b$ decrease with
growing $z$, it is also possible to have a scenario where
one of the scale factor shrinks, \ie $kl<0$.
The requirement that the zero-mode localizes translates in
this case ($kl<0$) into a condition for the cosmological constants:
\be{lambconst}
{3\over 5}\Lambda_{\theta}<\Lambda_z<-{1\over 5}\Lambda_{\theta},
\ee
where the first inequality comes from $kl<0$ whereas
the second inequality comes from $2k+l>0$.
This means that we can have an exponentially growing
infinite $z$-dimension as long as the $\theta$-dimension decreases rapidly
enough. The requirement for the localization of the
zero-mode also constrains the allowed values of brane tension,
via (\ref{statjumps}), $4\sigma>\sigma_{\theta}$.

\subsection{Massive RS-modes}

In the 6D scenario with 5D branes the bulk particles may propagate towards two
perpendicular directions. They can propagate in the compact, locally
flat dimension labelled by $\theta$ or towards the radial direction
labelled by $z$. Thus we have pure KK-excitation modes in both directions
together with mixed modes. Because the excitation along the compact 
dimension resembles clearly the ones in ADD-models, we call them ADD-modes.
In the very same spirit we call the perpendicular excitations as RS-modes,
because there share common properties with excitation of RS-models.
Again, note that the SM particles also have the ADD-modes but not the RS-ones. 
We study next the massive RS-modes and after that the ADD-modes.
Similar considerations have been carried out in \cite{misha,ross}.

Assuming the exponential forms (\ref{asol1})-(\ref{bsol1})
for the scale factors $a$ and $b$ (with $z_0=z_1$), 
(\ref{psiequ}) takes the form
\be{psiequ2}
\psi''-(l+4k)\psi'+\Big({m^2\over a_0^2}-
{n^2\over b_0^2}e^{2(l-k)(z-z_0)}\Big)e^{2k(z-z_0)}\psi=0.
\ee
We can transform this into the form of a 2D Schr\"odinger equation
by writing
\be{trans}
\rho={1\over k}\Big(\e^{k(z-z_0)}-1\Big),\ \; \psi=(k\rho+1)^{3\over 2}\Psi,
\ee
in which case (\ref{psiequ2}) transforms into
\be{shcr}
-{1\over 2b}\partial_{\rho}(b\ \partial_{\rho}\Psi)+\Big({3\over 8}{2l+5k
\over (k\rho+1)^2}+
{n^2\over 2b_0^2}(k\rho+1)^{2(l/k-1)}\Big)\Psi={m^2\over 2a_0^2}\Psi.
\ee
From (\ref{shcr}) we see that the angular modes $n\neq 0$ are 
separated by a mass gap $a_0/b_0$ from the
zero-mode on the brane.
The $s$-waves can be solved from (\ref{psiequ}):
\be{swave}
\psi_m(z)=e^{{l+4k\over 2}(z-z_0)}C_m[Y_{1+{l\over 2k}}({m\over a_0 k})
J_{2+{l\over 2k}}({m\over a_0 k}e^{k(z-z_0)})-
J_{1+{l\over 2k}}({m\over a_0 k}Y_{2+{l\over 2k}}(
{m\over a_0 k}e^{k(z-z_0)}))],
\ee
where $C_m$ is determined by the normalization condition
\be{psimnorm}
\int dz\, a^2b\, \psi_m\psi_n=\delta_{mn}.
\ee
In the case of an infinite extra-dimensions, 
the mass spectrum is continuous, starting from $m=0$ \cite{misha}.
The correction to the Newton's law due to the $s$-modes 
is \cite{ross,csaki}
\be{newtcorr}
\Delta V(r)\sim r^{-3-{l\over k}}\equiv r^{-\alpha}.
\ee
The exponent is bounded due to the requirement of
zero-mode normalizability, $2k+l>0$. Hence, we see that
the power of the correction (\ref{newtcorr}) gives us 
information of the internal structure of the extra
dimensions: if $1<\alpha<3$, one of the scale factors
is growing while the other is shrinking. Note that 
this also indicates the presence of more than
one extra dimension since in the RS-scenario $\alpha=3$.
If on the other hand, $\alpha\gsim 3$, both
scale factor are decreasing with $z$.

Putting another brane at $z=z_1$, Fig. 1c,
modifies the normalization of the graviton 
wave-functions, $\psi_m$, as well as discretizes the
mass spectrum of the gravitons. The $s$-waves (\ref{swave})
are such that $\psi'(z_0)=0$. When we now
set another brane at $z=z_1$, we must again require that
$\psi'(z_1)=0$, which obviously then leads to a discretization
of the allowed graviton masses, $m$. The requirement $\psi'(z_1)=0$
leads to a condition
\be{discm}
J_{1+{l\over 2k}}({m\over k a_0}e^{k(z_1-z_0)})=
J_{1+{l\over 2k}}({m\over k a_0}),
\ee
which needs to be solved numerically to obtain the
exact mass spectrum. However, we can estimate the mass gap
between the states easily for large $e^{k(z_1-z_0)}$,
$\Delta m\approx \pi k a_0 e^{-k(z_1-z_0)}$. 
Note that the expression for the mass gap is valid only at large
$m$.

\subsection{ADD-modes}

Since in this scenario we also have a compact dimension, ADD-gravitons
propagating along the brane are present. The radius of the compact dimension is
given by the ratio of the scale factor at the brane at $z=z_0$,
\be{comprad}
R(z_0)={b(z_0)\over a(z_0)}.
\ee
The spectrum of ADD-masses can be read from (\ref{psiequ}),
\be{kkmasses}
m_n={a_0\over b_0}n.
\ee
It is crucial to note, that this is the mass formula also for particles 
localized at the 4-brane. In particular the standard model particles have 
excitation given by (\ref{kkmasses}). This put limits on the 
mass of the lightest massive excitation $m_{ADD}={a_0\over b_0}$.
Using the expressions (\ref{asol1}), (\ref{bsol1}) we get
\be{comrad2}
R(z)={b_0\over a_0}e^{(k-l)(z-z_0)},\, m_n(z)={a_0\over b_0}
e^{(l-k)(z-z_0)}n,
\ee
from which we see that the ADD-spectrum on different branes
can vary greatly. Moreover,
the correction to the Newton's law due to ADD-spectrum in well known
\cite{arka}.
It can be shown that the leading correction to the Newton's law is
$\Delta V(r)\sim r^{-2}$ because the compact dimension is 1-dimensional.

\subsection{4D gravitational constant and corrections to
Newtonian potential}

Since the gravitational action is of the form
\be{gravact}
S_{grav}={1\over 16\pi G_6}\int d^4x\int d\theta\, dz\ \sqrt{g_{6}}\cR_6+...,
\ee
we see that the 4D Newton's constant in a scenario with two 5D branes,
is given by
\be{4dnc}
G_4={2k+l\over 2\pi a_0^2b_0 [1-e^{-(2k+l)(z_1-z_0)}]}G_6.
\ee
If we normalize the scale factors
so that on the other brane $a^2(z_1)b(z_1)=1$, we get instead
\be{4dnd}
G_4={2k+l\over 2\pi [e^{(2k+l)(z_1-z_0)}-1]}G_6.
\ee

In the six dimensional scenario, unlike in the RS-scenarios,
one can accommodate a setup where gravity on the positive 
tension brane is suppressed. As an example, consider a
two brane setup with branes at $z=0$ and $z=z_1$. From 
(\ref{statjumps}) we see that the brane at the origin has a
positive brane tension if $l+3k>0$. 
On the other hand, from (\ref{4dnc}) it is clear that
if we wish to alleviate the hierarchy problem on the same brane,
$2k+l<0$. Hence, $k$ and $l$ must satisfy the conditions
\be{nicestate}
-{l\over 3}<k<-{l\over 2},\ k>0,\ l<0.
\ee
Condition (\ref{nicestate}) implies that $b(z)$ grows
while $a(z)$ shrinks with increasing $z$. Such a setup is not
possible in the normal RS-scenarios and becomes possible only in
models with more than one extra dimension \cite{leblond}.

We can now also compare the corrections to the Newtonian potential arising 
from the RS- and ADD-excitations. We have already noted that the sign of the
product $kl$ reflects to the power of the leading RS-correction.
If we compare the leading ADD-correction $\sim r^{-2}$ and the RS-correction
$\sim r^{-3-{l\over k}}$, we find that if $l/k < -1$, the RS-correction is
the leading one. In the opposite case ADD-correction is the leading one.
However, one must keep in mind that in realistic models the ADD-correction 
is only significant at distances smaller than the radius of the compact 
extra dimension,
\ie $r\lsim 1/\TeV$, and cannot be measured directly. One can then
expect that measurements on the short range gravitational potential can 
reveal structure about extra dimensions due to the RS-excitations.
Hence, we can gain information from the very small compact
dimensions, even though they are too small to be studied directly,
by studying the RS-excitations.


\section{Bowl-universe}
We have already seen that there exists at least two types of solutions
to the Einstein's equations in 6D,
the trivial constant solution as well as the 
exponential solutions. With branes added to the picture, 
we can construct a large number of different types of scenarios, some
of which were shown in Section \ref{branesetups}.
In the case of exponentially growing or shrinking scale factors,
it is obvious that if we wish to include the origin, $z=0$, 
in our space, we must place a string-like brane at the origin
(unless the 6D space around the origin is flat). It is therefore 
interesting to ask whether this is always necessary or is there a solution
which does not require any matter at the origin.

If there is no matter at the origin, the second derivative of the
scale factors must be continuous. Hence, we must require that
$a'(0)=0,\ b'(0)=0$. From the second Einstein's equation, Eq. (\ref{ste2}),
we immediately see that $\Lambda_z=0$. From the set of static 
Einstein's equations we find that there are three different 
solutions which smoothly include the origin:
\bea{piesols}
\left . \begin{array}{cccc}
\textrm{I}: & a(z)=a_0 & b(z)=b_0 & \Lambda=\Lambda_{\theta}=0\\
\textrm{II}: & a(z)=a_0 & b(z)=b_0{\cosh(kz)\over\cosh(kz_0)} &
 k^2=-\Lambda,\ \Lambda_{\theta}=0\\
\textrm{III}: & a(z)=a_0{\cosh^{2/5}(\kappa z)\over\cosh^{2/5}(\kappa z_0)} & 
b(z)=b_0{\textrm{sech}^{3/5}(\kappa z)\over\textrm{sech}^{3/5}(\kappa z_0)} & 
k^2=-\Lambda,\ \Lambda_{\theta}={8\over 3}\Lambda,\\
\end{array} \right .
\eea
where $\kappa=k/\sqrt{10}$.
Note that the sign of $\Lambda$ is very significant in the constructions:
if $\Lambda>0$, $z$ is bounded by the requirement $a(z)>0,\ b(z)>0$, and
hence
\bea{piezbound}
\textrm{II}: & z < & {\pi\over 2\sqrt{\Lambda}}\non\\
\textrm{III}: & z < & \pi\sqrt{2\over 5\Lambda}\non
\eea
Note also, that this type single brane solution has no 5-D brane analogy!
However, there is a singularity at finite distance from the brane.

We now add a brane at $z=z_0$ to the setup and 
consider a case where the brane has been wrapped around the
origin so that the SM fields can propagate along the compact
extra dimension. This opens an interesting possibility
that there is no
space outside the brane, \ie gravitons can only propagate inwards
as well as along the brane. In the cases II and III, the
volume element $a^3 b$ is a bowl-shaped function and therefore 
we refer to this particular model as the bowl-model. 
Since SM fields and ADD-gravitons are present in the particle
spectrum, the size of the compact extra dimension must be small
enough so that the lightest ADD-excitations are out of reach of the
collider experiments, \ie $m_{ADD}\gsim 1 \TeV$.

The tension of the single 4-brane in the different scenarios
is easily calculable from (\ref{statjumps}). In the trivial 
first case we see that there can be no matter on the brane 
unless we allow for a non-trivial space-time structure outside 
the space bounded by the brane.
The second and the third case are more interesting: In the second case 
we see that $\sigma$ is always positive while $\sigma_{\theta}$ vanishes. 
In the third case, $\sigma$ is positive for $\cosh(\kappa z_0)>2/3$
and $\sigma_{\theta}>0$ for all $z_0$.

The value of the gravitational constant on the brane is easily
calculable from (\ref{gravact}):
\be{gravconst}
{1\over 2\pi a_0^2b_0}{G_6\over G_4}=\left\{\begin{array}{cc}
z_0, & (\I)\\
\cosh^{-1}(kz_0)\Big({z_0\over 2}+{1\over 4k}\sinh(2kz_0)\Big) & (\II)\\
\cosh^{-1/5}(\kappa z_0)\int_0^{z_0}\cosh^{1/5}(\kappa z), & (\III)
\end{array}\right .
\ee
where the integral in the third case can be  approximated by
\be{iiiapp}
\int_0^{z_0}\cosh^{1/5}(\kappa z)\approx
\left\{\begin{array}{cc}
z_0+{5\over 12}\kappa^2z_0^3, & \kappa z_0<1\\
{31\over 75}{1\over\kappa}+2^{4/5}{1\over\kappa}(
e^{\kappa z_0/2}-e^{1/5}), & \kappa z_0>1.
\end{array}\right .
\ee

\subsection{Gravitons}
The graviton spectrum is again given by Eq. (\ref{psiequ}).
In all of the cases, a massless zero-mode $\psi=\psi_0$
is present. The existence of massive modes is dependent
on the behaviour of the scale factors.

\subsubsection{Case I}
The trivial solution, $a=a_0$ and $b=b_0$, is just a
scenario with two compact flat dimensions. In our setup,
the SM fields feel one of the extra dimensions while
gravitons can propagate along both of them. Clearly such a setup
is insufficient in alleviating the hierarchy problem.

\subsubsection{Case II}
Consider the graviton equation in case II. A zero-mode, $\psi=\psi_0$,
is present as always. The massive graviton modes are described by
\be{pieIIgrav}
\psi''+k \textrm{tanh}(kz)\psi'+\Big({m^2\over a_0^2}-{n^2\over b_0^2}
{\cosh^2(kz_0)\over\cosh^2(kz)}\Big)\psi=0,
\ee
which has a solution with $\psi'(0)=0$, of the form ($m=0$)
\be{pieIIgravsol}
\psi(z)=C_m \Big(P_{\mu}(i\sinh(kz))Q_{\mu-1}(0)-Q_{\mu}(i\sinh(kz))P_{\mu-1}(0)\Big),
\ee
where 
\be{mu} 
\mu={1\over 2}\Big(-1+\sqrt{1-4{m^2\over k^2a_0^2}}\Big)
\ee
and $P$ and $Q$ are the Legendre functions. The overall constant factor
is determined by the normalization condition, (\ref{psimnorm}).
The spectrum of the graviton masses is strongly dependent on the
sign on $\Lambda$, just like the behaviour of the scale factors.
If $\Lambda>0$, there are no massive $s$-waves with $\psi'(z_0)=0$.
On the other hand, higher waves, with $n>0$, do exist.

If, however $\Lambda<0$, $s$-waves are present in the graviton
spectrum. To find the graviton masses, 
we require that  $\psi'(z_0)=0$, \ie the derivative vanishes on the brane.
This is in general a numerical problem. From the numerical work
it is clear that as $z_0$ grows, the mass spectrum becomes more dense
and the mass of the lightest massive mode decreases.

\subsubsection{Case III}
In the third case, the graviton spectrum is calculable from 
\be{pieIIIgrav}
\psi''+\kappa\textrm{tanh}(\kappa z)\psi'+\Big({m^2\over a_0^2}
{\cosh^{4/5}(\kappa z_0)\over \cosh^{4/5}(\kappa z)}-{n^2\over b_0^2}
{\cosh^{6/5}(\kappa z)\over\cosh^{6/5}(\kappa z_0)}\Big)\psi=0.
\ee
Eq. (\ref{pieIIIgrav}) needs to be solved numerically.
Again, $\Delta m$ and the mass of the lightest mode decrease
with increasing $z_0$.

In the large $\sinh(\kappa z)$ limit, $s$-waves can be approximated by
\be{pieIIIapp}
\psi(z)=\sinh^{-1/2}(\kappa z)[A_mJ_{5/4}\Big({5m\over 2k\tilde{a}_0}
\sinh^{-2/5}(\kappa z)\Big)+B_mY_{5/4}\Big({5m\over 2k\tilde{a}_0}
\sinh^{-2/5}(\kappa z)\Big)],
\ee
where $\tilde{a}_0=a_0/\cosh^{2/5}(\kappa z_0)$.

\subsection{Hierarchy problem}
We can now look for parameters which alleviate the hierarchy
problem while keeping the KK-excitations along the brane
heavy. Assume that the fundamental 6D gravity scale is $M_*$.
The non-observation of KK-excitations combined with (\ref{gravconst}) then
gives the following constraints for the value of the parameters:
\be{6dmass}\begin{array}{cccccc}
{1\over 2\pi a_0^3}m_{ADD}z_0^{-1}\MP^2 & \lsim & M_*^4 & \lsim &
{1\over 2\pi b_0^3}z_0^{-1}{\MP^2\over m_{ADD}^2} & (\I)\\
{2k\over\pi a_0^3}e^{-kz_0}m_{ADD}\MP^2 & \lsim & M_*^4& \lsim & 
{2k\over\pi b_0^3}e^{-kz_0}{\MP^2\over m_{ADD}^2} & (\II)\\
{\kappa\over 4\pi a_0^3}e^{-3\kappa z_0/10}m_{ADD}\MP^2 & \lsim & M_*^4& \lsim &
{\kappa\over 4\pi b_0^3}e^{-3\kappa z_0/10}{\MP^2\over m_{ADD}^2}, & (\III)
\end{array}
\ee
where the experimental limit for the mass of the lightest ADD-mode is 
denoted by $m_{ADD}$. We are interested in the lower limit
for $M_*$. Assuming that $\tilde{m}=1\TeV$ and using $a_0=1$, we get
\be{6dnumass}
\begin{array}{cccc}
(z_0/\GeV^{-1})^{-1/4}\times 10^{10}\GeV & \lsim & M_* & (\I)\\
(k/\GeV)^{1/4}e^{-kz_0/4}\times 10^{10}\GeV & \lsim & M_* & (\II)\\
(\kappa/\GeV)^{1/4}e^{-3 \kappa z_0/40}\times 10^{10}\GeV& \lsim & M_* & (\III).
\end{array}
\ee
If we wish that the fundamental 6D gravity scale is $\sim 1\TeV$,
we must in the I case require that $z_0\sim 10^{28}\GeV^{-1}$, making
the hierarchy problem worse by introducing a new, large scale to the theory.
In case II, we see that if $k\sim M_*$, $kz_0\sim 90$, where as in case III, 
$\kappa z_0\sim 300$. A setup with the fundamental scale around $1\TeV$ is hence
feasible in this scenario. The graviton mass spectrum in this is case
is well approximated by a continuous spectrum, starting at $m=0$.


\section{Cosmology}

In addition to the interesting properties that static brane world 
configurations possess, the dynamical evolution of space-time also offer
novel properties that have an effect on cosmology. Cosmological evolution 
in the brane world scenarios have been shown to possess
interesting properties \cite{bdl}-\cite{ida}. In order to study cosmology on a
brane in the 6D scenario, we adopt a metric:
\be{cosmometric}
ds^2=\eta(\tau,z)^2d\tau^2-R(\tau,z)^2\delta_{ij}dx^idx^j-a(\tau,z)^2dz^2-
b(\tau,z)^2d\theta^2.
\ee
The components of the Einstein's tensor $G_{\mu\nu}$ are:
\bea{cosmoes}
G_{00} & = & - \frac{{{\eta}}^2\,a'\,b'}
     {{a}^3\,b}  - 
  \frac{3\,{{\eta}}^2\,a'\,R'}
   {{a}^3\,R} + \frac{3\,{{\eta}}^2\,b'\,
     R'}{{a}^2\,b\,R} + 
  \frac{3\,{{\eta}}^2\,{R'}^2}
   {{a}^2\,{R}^2} + \frac{{{\eta}}^2\,b''}
   {{a}^2\,b} +\\
& &  \frac{3\,{{\eta}}^2\,R''}
   {{a}^2\,R} - \frac{\dot{a}\,\dot{b}}
   {a\,b} - \frac{3\,\dot{a}\,\dot{R}}{a\,R} - 
  \frac{3\,\dot{b}\,\dot{R}}{b\,R} - 
  \frac{3\,{\dot{R}}^2}{{R}^2}\\
G_{11} & = & \frac{{R}^2\,a'\,b'}{{a}^3\,b} + 
  \frac{{R}^2\,a'\,{\eta}'}
   {{a}^3\,{\eta}} - 
  \frac{{R}^2\,b'\,{\eta}'}
   {{a}^2\,b\,{\eta}} + 
  \frac{2\,R\,a'\,R'}{{a}^3} - 
  \frac{2\,R\,b'\,R'}{{a}^2\,b} - 
  \frac{2\,R\,{\eta}'\,R'}
   {{a}^2\,{\eta}} - 
  \frac{{R'}^2}{{a}^2} -\\ 
& &   \frac{{R}^2\,b''}{{a}^2\,b} - 
  \frac{{R}^2\,{\eta}''}
   {{a}^2\,{\eta}} - 
  \frac{2\,R\,R''}{{a}^2} + 
  \frac{{R}^2\,\dot{a}\,\dot{b}}
   {a\,b\,{{\eta}}^2} - 
  \frac{{R}^2\,\dot{a}\,\dot{\eta}}
   {a\,{{\eta}}^3} -
  \frac{{R}^2\,\dot{b}\,\dot{\eta}}
   {b\,{{\eta}}^3} +\\ 
& &   \frac{2\,R\,\dot{a}\,\dot{R}}
   {a\,{{\eta}}^2} + 
  \frac{2\,R\,\dot{b}\,\dot{R}}
   {b\,{{\eta}}^2} - 
  \frac{2\,R\,\dot{\eta}\,\dot{R}}
   {{{\eta}}^3} + 
  \frac{{\dot{R}}^2}{{{\eta}}^2} + 
  \frac{{R}^2\,\ddot{a}}{a\,{{\eta}}^2} + 
  \frac{{R}^2\,\ddot{b}}{b\,{{\eta}}^2} + 
  \frac{2\,R\,\ddot{R}}{{{\eta}}^2}\\
G_{44} & = &  - \frac{b'\,{\eta}'}
     {b\,{\eta}}  - 
  \frac{3\,b'\,R'}{b\,R} - 
  \frac{3\,{\eta}'\,R'}
   {{\eta}\,R} - 
  \frac{3\,{R'}^2}{{R}^2} - 
  \frac{{a}^2\,\dot{b}\,\dot{\eta}}
   {b\,{{\eta}}^3} + 
  \frac{3\,{a}^2\,\dot{b}\,\dot{R}}
   {b\,{{\eta}}^2\,R} -\\ 
& &   \frac{3\,{a}^2\,\dot{\eta}\,\dot{R}}
   {{{\eta}}^3\,R} + 
  \frac{3\,{a}^2\,{\dot{R}}^2}
   {{{\eta}}^2\,{R}^2} + 
  \frac{{a}^2\,\ddot{b}}{b\,{{\eta}}^2} + 
  \frac{3\,{a}^2\,\ddot{R}}{{{\eta}}^2\,R}\\
G_{55} & = &  \frac{{b}^2\,a'\,{\eta}'}
   {{a}^3\,{\eta}} + 
  \frac{3\,{b}^2\,a'\,R'}{{a}^3\,R} - 
  \frac{3\,{b}^2\,{\eta}'\,R'}
   {{a}^2\,{\eta}\,R} - 
  \frac{3\,{b}^2\,{R'}^2}{{a}^2\,{R}^2} - 
  \frac{{b}^2\,{\eta}''}
   {{a}^2\,{\eta}} - 
  \frac{3\,{b}^2\,R''}{{a}^2\,R} -\\ 
& &   \frac{{b}^2\,\dot{a}\,\dot{\eta}}
   {a\,{{\eta}}^3} + 
  \frac{3\,{b}^2\,\dot{a}\,\dot{R}}
   {a\,{{\eta}}^2\,R} - 
  \frac{3\,{b}^2\,\dot{\eta}\,\dot{R}}
   {{{\eta}}^3\,R} + 
  \frac{3\,{b}^2\,{\dot{R}}^2}
   {{{\eta}}^2\,{R}^2} + 
  \frac{{b}^2\,\ddot{a}}{a\,{{\eta}}^2} + 
  \frac{3\,{b}^2\,\ddot{R}}{{{\eta}}^2\,R}\\
G_{40} & = & - \frac{b'\,\dot{a}}{a\,b}   - 
  \frac{3\,R'\,\dot{a}}{a\,R} - 
  \frac{{\eta}'\,\dot{b}}
   {b\,{\eta}} - 
  \frac{3\,{\eta}'\,\dot{R}}
   {{\eta}\,R} + \frac{b^{(1,1)}}{b} + 
  \frac{3\,\dot{R'}}{R}
\eea
Matter on brane has the form
\be{cosmobrane}
T_A^B=f(\tau,z)\diag(\rho(\tau),-P(\tau),-P(\tau),-P(\tau),0,-P_v(\tau)),
\ee
with an arbitrary prefactor $f(\tau,z)$, whose significance becomes
clear later on.
Jumps are required for components (0,0), (1,1), (5,5):
\bea{cosmojumps}
{1\over a_0^2}{[b']\over b}\bigg |_{z_0}+3{1\over a_0^2}{[R']\over R}\bigg |_{z_0}
& = & 8\pi G_6f_0\rho\nonumber\\
{1\over a_0^2}{[b']\over b}\bigg |_{z_0}+{1\over a_0^2}{[\eta']\over \eta}\bigg |_{z_0}+
2{1\over a_0^2}{[R']\over R}\bigg |_{z_0} & = & -8\pi G_6f_0 P\\
{1\over a_0^2}{[\eta']\over \eta}\bigg |_{z_0}+3{1\over a_0^2}{[R']\over R}\bigg |_{z_0}
& = & -8\pi G_6f_0P_v\nonumber,
\eea 
where the index $0$ refers to values on the brane.
The jumps are in the direction perpendicular to the brane
and can easily be calculated from the Gauss-Codacci -equations
with the unit vector field normal to the 4-brane chosen as
$n^A=(0,0,0,0,1/a,0)$.

The non-trivial continuity equations (${T_A^B}_{;B}=0$) are:
\bea{cosmocont} 
{T_0^B}_{;B} & = & \dot{\rho}+ 
  3(P+\rho)\frac{\dot{R}}{R} + 
\rho(\frac{\dot{a}}{a} + 
  \frac{\dot{b}}{b} + 
  \frac{\dot{f}}{f}) + 
  P_v \frac{\dot{b}}{{b}}=0\\
{T_4^B}_{;B} & = & P_v\frac{b'}
{{b}} - \rho\frac{\eta'}{\eta} + 
  3P\frac{R'}{R}=0.
\eea
From the ${T_0^B}_{;B}=0$ continuity equation we see that in order
to recover the usual continuity equation on the brane we have a number
of different choices:
\be{fchoice}
\begin{array}{ccc}
(i) & \dot{b}=0 & f=1/a\\
(ii) & P_v=0 & f=1/(ab)\\
(iii) & P_v=-\rho & f=1/a.
\end{array}
\ee
In each case we can then solve for the jump factors from (\ref{cosmojumps}).
From the first continuity equation we see that the dynamics of the extra
dimensions can also lead to the non-conservation of energy density on the
brane.

The jump of (4,4) gives a constraint relating the mean values of the
scale factors,
\be{cjump44}
\rho {\sharp\eta'\sharp\over\eta_0}-3 P {\sharp R'\sharp\over R_0}-
P_v {\sharp b'\sharp\over b_0}=0,
\ee 
where $\sharp f\sharp\equiv {1\over 2}(f(z_0+)+f(z_0-))$ is the mean value
across the brane. Note that is also apparent from the (4) 
component of the continuity equations.

Taking the mean value of $G_{44}$ and 
assuming for simplicity that space on the brane is locally 
invariant under space inversions about the brane, 
\ie $\sharp R'\sharp=0$, and choosing $\eta_0=1$,
we get the Friedmann-type evolution equation of the scale factor 
$R$ on the brane:
\be{cmean44v2}
  \frac{\ddot{R}}{R}+
  (\frac{{\dot{R}}}{{R}})^2+
  \frac{\dot{b}}{b}{\dot{R}\over R}=
-{8\pi G_6f_0\over 32}\Big((\rho+P-P_v)^2+3\rho P_v\Big)+{1\over 3}
({\sharp b'\sharp\over a b})^2{P_v\over\rho}
-\frac{\ddot{b}}{3 b}+{1\over 3}\Lambda_z.
\ee
We see that the evolution of the scale factor is, like in the
RS-scenario \cite{bdl}, fundamentally different from the Friedmann
equation since here $H=\dot{R}/R\sim\rho$, instead of $H^2\sim\rho$.
This is in fact a general property of extra dimensional models with
a single warped extra dimension, as one can see from considering a
$D$ dimensional metric,
\be{dmetric} 
ds^2=\eta(\tau,z)^2d\tau^2-R(\tau,z)^2\delta_{ij}dx^idx^j-a
(\tau,z)^2dz^2-\Sigma^{D-5}_ib_i(\tau,z)^2d\theta_i^2,
\ee
along with matter on the brane, $T_A^B=f(\tau,z)\, \diag(\rho,-P,-P,-P,0,-P_1,...,
-P_{D-5})$.
From the Gauss-Codacci -equations one can easily show that in
the case of a (locally) spatially symmetric brane, the four dimensional
curvature scalar (and hence $H^2$) is proportional to 
$T^2/(D-2)-T_{AB}T^{AB}$ and one cannot choose $P_i\in\real$ in such a way
that the $\rho^2$ term vanishes.
One can recover the standard Friedmann equation on the brane
by adding a energy density on the brane, like in the RS-scenario 
\cite{cscosmo,cline}.

From (\ref{cmean44v2}) we also see that in the 6D brane world the evolution 
of the compact dimension can significantly affect the evolution
on the brane. For example, if $\dot{b}/b$ is large compared to
$H$, we have situation where $H\sim\rho^2$. The evolution of the 
scale factor can hence be much more complicated than in
standard cosmology. Furthermore, a changing $b$ also would indicate
a varying tower of KK-masses as well as have an effect on the
gravitational constant on the brane.


\section{Conclusions}

In the present paper we have studied the possibility of having a six dimensional
theory with a 5-brane such that one dimension of the brane forms a small 
compact space. In this scenario we thus have one ADD-like spatial dimension
together with one Randall-Sundrum -like dimension. The size of the ADD-space
is naturally constrained by the requirement that the Kaluza-Klein excitations 
are heavy enough compared to the experimental limits. Besides our brane where 
standard model particles live, these models may have an additional brane which
can be 4-dimensional but may also be 5-dimensional. As a special case, there 
is the natural model with only a single 3-brane \cite{misha}. Obviously, 
one can also extend the considerations presented here to scenarios with 
several compact dimensions along with a single warped dimension. 

Six dimensional constructions make possible aspects that one cannot
have in 5D models. An interesting property is the possibility that
one can have a positive tension brane while at the same time
alleviating the hierarchy problem. This is made possible by the extra degree 
of freedom introduced by the scale factor of the compact dimension.

An interesting case among numerous other possible constructions, is a simple
model with only one 5-dimensional brane. In this single brane model the 
parameters are chosen so that there is no extra brane at the origin, 
something one cannot do in 5D construction. This bowl-model has 
thus no mirror world and it includes the interesting property
that the hierarchy problem is solved with a single positive tension brane.
As we have seen, one can construct both one and two brane models where 
ordinary matter lies on a positive tension brane while solving the 
hierarchy problem. Phenomenologically these models differ from each other;
\eg there can be both gravitational and SM neutrino interactions between 
two branes, which are obviously absent in the single brane scenario.
These pleasant properties of the single brane construction make it interesting among 
6D models. The bowl-model, and other 6D constructions, can also 
in principle be tested experimentally by observing the type of the 
leading correction to the Newtonian gravitational potential. The power
of the correction can reveal information of the structure
of the extra dimensions and possibly distinguish between different 
types of models, even if the radius of the compact dimension is
much too small to be detected directly.

Cosmologically, these 6-dimensional models, thus including the bowl-model,
share the difficulties of all extra dimensional models on restoring 
consistently the ordinary Friedmann evolution on the brane 
\cite{cscosmo,cline}. Thus, a more detailed study of
the cosmological models is clearly needed.

It would be also interesting to study more some of the other features 
of the 6D models, and in particular the simple bowl-model. 
Issues, like creation of the matter by dynamics certainly would
have interest of its own. Also the possibility to have a changing 
gravitational
constant and ADD-excitation masses, {\it i.e.} graviton and standard model 
excitation masses, would certainly have interesting consequences.
The bowl-model also allows a construction where with a positive cosmological 
constant together with 6-dimensional brane there is at a finite coordinate 
distance a singularity like in some Randall-Sundrum -models. The effect of 
these possible features to cosmology remains to be studied.
Stabilization of the extra dimensions is also an important issue that
needs further consideration \cite{burgess,kanti}.

Six dimensional models with a compact and a warped dimension open
novel perspectives to the brane world models. It seems that one can 
combine nice features of both ADD- and RS-models in such a way 
that many of the central problems are alleviated. 


\section*{Acknowledgments}
This work has been partly supported by the Magnus Ehrnrooth
Foundation.



\begin{thebibliography}{X}
\bibitem{add} N. Arkani-Hamed, S. Dimopoulos and G. Dvali, \pl{B429}{1998}{263};
I. Antoniadis \etal, \pl{B436}{1998}{257}.
\bibitem{rs} L. Randall and R. Sundrum, \prl{83}{1999}{3370}; L. Randall and R. Sundrum, \prl{83}{1999}{4690}.
\bibitem{rubakov} {\it For a review, see} \eg V. A. Rubakov, 
{\it Phys. Usp.} {\bf 44} (2001) 871-893; {\it Usp. Fiz. Nauk} {\bf 171} (2001) 913-938 (hep-ph/0104152).
\bibitem{misha} T. Gherghetta and M. Shaposhnikov, \prl{85}{2000}{240}.
\bibitem{olive} P. Kanti, R. Madden and K. Olive, \pr{D64}{2001}{044021}.

\bibitem{ross} I. Kogan \etal, \pr{D64}{2001}{124014}.
\bibitem{chacko} Z. Chacko \etal, \jhep{0203}{2002}{001}; Z. Chacko and A. E. Nelson, \prd{62}{2000}{085006}.
\bibitem{casi} I. Kogan and N. Voronov, \jetpl{38}{1983}{311};
P. Candelas and S. Weinberg, \np{B237}{1984}{397}.
\bibitem{csaki} C. Csaki \etal, \np{B581}{2000}{309}.
\bibitem{arka} N. Arkani-Hamed, S. Dimopoulos and G. Dvali, \pr{D59}{1999}{086004}. 
\bibitem{leblond} F. Leblond, R. C. Myers and D. J. Winters, \jhep{0107}{2001}{031}.
\bibitem{bdl} P. Bin\'etruy, C. Deffayet and D. Langlois, \np{B565}{2000}{269}.
\bibitem{cscosmo} C. Csaki \etal, \pl{B462}{1999}{34}.
\bibitem{cline} J. Cline, C. Grojean and G. Servant, \prl{83}{1999}{4245}.
\bibitem{ida} D. Ida, \jhep{0009}{2000}{014}.
\bibitem{burgess} C. P. Burgess \etal, \jhep{0201}{2002}{014}.
\bibitem{kanti}  P. Kanti, K. Olive and M. Pospelov, \pl{B538}{2002}{146}.
\end{thebibliography}
\end{document}